\documentclass[11pt]{article}
\usepackage{amsmath,amssymb}
\usepackage{mathrsfs}

\usepackage{graphicx}
\usepackage{color}
\usepackage{caption}
\usepackage{float}
\usepackage{calligra}
\usepackage[
colorlinks=true,
linkcolor=blue,
urlcolor=blue,
filecolor=black,
citecolor=red,
]{hyperref}
\usepackage{tikz}
\usepackage{tikz-cd}
\usetikzlibrary{babel,cd}
\usetikzlibrary{calc, arrows, decorations.markings}
\def\eg{{\it e.g. }}
\def\ie{{\it i.e.}}

\def\Tr{\mathrm{Tr}\,}
\def\p{\partial}
\def\bp{\bar{\partial}}
\def\la{\langle}
\def\ra{\rangle}

\newcommand{\be}{\begin{equation}}
\newcommand{\ee}{\end{equation}}

\newcommand{\bw}{\bar{w}}
\newcommand{\bz}{\bar{z}}
\newcommand{\bx}{\bar{x}}
\newcommand{\by}{\bar{y}}

\newcommand{\CM}{{\mathcal{M}}}
\newcommand{\CA}{{\mathcal{A}}}
\newcommand{\CF}{{\mathcal{F}}}
\newcommand{\CO}{{\mathcal{O}}}

\newcommand{\CK}{{\mathcal{K}}}

\newcommand{\CD}{{\mathcal{D}}}

\usepackage{calligra}

\DeclareMathAlphabet{\mathcalligra}{T1}{calligra}{m}{n}
\DeclareFontShape{T1}{calligra}{m}{n}{<->s*[2.2]callig15}{}

\newcommand{\bea}{\begin{eqnarray}}
\newcommand{\eea}{\end{eqnarray}}
\newcommand{\ba}{\begin{array}}
\newcommand{\ea}{\end{array}}

\newcommand{\BC}{{\mathbb{C}}}

\newcommand{\CC}{\mathcal{C}}
\newcommand{\CR}{\mathcal{R}}
\newcommand{\CE}{\mathcal{E}}
\newcommand{\bT}{\bar{T}}
\begin{document}
\title{\textbf{Energy of Decomposition and
Entanglement Thermodynamics for $T^2$-deformation}}

	\vspace{2cm}
	\author{ \textbf{Kuroush Allameh and  Amin Faraji Astaneh}} 
	\date{}
	\maketitle
	\begin{center}
		\hspace{-0mm}
		\emph{ Department of Physics, Sharif University of Technology,\\
			P.O.Box 11155-9161, Tehran, Iran}
	\end{center}
	


	\begin{abstract}
		\noindent{
We have presented a set of laws of entanglement thermodynamics for $T\bar{T}$-deformed CFTs and in general for $T^2$-deformed field theories. In particular, the first law of this set, states that although we are dealing with a non-trivial deformed theory, the change of the entanglement entropy is simply translated to the change of the bending energy of the entangling surface. We interpret this energy as the energy of decomposition. Probing the whole spectrum of the deformed theory, a second law also results, which suggests an inequality that the first law is derived from its saturation limit. We explain that this second law guarantees the preservation of the unitarity bound. The thermodynamical form of these laws requires us to define the temperature of deformation and express its characteristics, which is the subject of the third law. We use a holographic approach in this analysis and in each case, we consider the generalization to higher dimensions.}
	\end{abstract}
	
	\vskip 8 cm
	\noindent
	\rule{7.7 cm}{.5 pt}\\
	\noindent 
	\noindent
	\noindent ~~~ {\footnotesize e-mails:\ k.allameh@physics.sharif.edu , faraji@sharif.ir }


	\newpage
\section{Introduction}
Entanglement Entropy (EE) is a very good measure of entanglement between the parts of a composite quantum mechanical system. These parts could be spins on cites of a lattice. Moving to the continuum limit we gradually arrive at a manifold on which a field theory lives. Then, partitioning the system into some subsystems makes geometrical sense. The calculation of EE depends on what state we calculate the entropy for, and how we isolate a subsystem.\\
At the leading orders and for the ground state of a local quantum field theory the EE admits an area law. The details of how to isolate a subsystem, appear in the subleading term which is some sort of a  \emph{Bending Energy} of the entangling surface.\footnote{Throughout this paper, we consider that the base manifold is flat and the time translational symmetry is guaranteed.}\\
Denoting the entangling surface by $\Sigma$, the EE takes the following expansion in UV cut off parameter, $\epsilon$
\be
S_{EE}=\frac{s_\CA}{\epsilon^{d-2}}\CA_\Sigma+\frac{s_\CE}{\epsilon^{d-4}}\CE_\Sigma+\cdots\, .
\ee
where $\CA_\Sigma$ is the area, and the bending energy is defined as the \emph{Willmore functional} of the entangling surface\footnote{More precisely the Willmore energy also includes the Euler characteristic of the surface such that $\CE=\int_\Sigma K^2-2\pi \chi_\Sigma$. However, since we only concentrate on a single class of topology, this constant could be disregarded without the loss of generality.}\cite{Willmore}-\cite{Astaneh:2014uba}.
\be
\CE_\Sigma=\frac{1}{8(d-2)^2}\int_\Sigma K^2\, ,
\ee
in which $K$ is the trace of the extrinsic curvature. The coefficients of the expansion take the following values
\be
s_\CA=\frac{2N^{\frac{d}{2}}}{(d-2)}\ \ , \ \ s_\CE=-\frac{(d-3)N^{\frac{d}{2}}}{(d-4)}\, ,
\ee
where $N$ is the effective number of degrees of freedom of the field theory.\footnote{We set $\frac{L^{d-1}}{G^{(d+1)}_N}=N^{d/2}$ where $L$ is the radius of AdS spacetime and $G_N$ is the Newton constant. This equality, in its exact form, includes some numerical factors as well, for example  $L/G_N^{(3)}=2/3N$ ($d=2$), $L^4/G_N^{(5)}=2/\pi N^2$ ($d=4$). We have absorbed these numbers in definition of $N$.}
This structure can be derived from the conformal anomaly as well as the holography. 

Our intuition leads us to call this energy the \emph{Energy of Decomposition}, which amounts to the energy that one needs to spend in order to isolate a subsystem of an entangled composite system. Intuitively, the more symmetrical our decomposition is, the less decomposition energy will be required. This statement is a retelling of Willmore's famous conjecture and proves that our definition of decomposition energy is a reliable one. Based on the Willmore's conjecture, there are global minimizers for the bending energy in each class of topology. For example for surfaces of genus zero, the sphere is the minimizer and for that
\be
\CE_{S^{d-2}}=\frac{\pi^\frac{d-1}{2}}{4\Gamma\left(\frac{d-1}{2}\right)}R^{d-4}\, .
\ee
For more explanation and investigation of other topology classes, refer to \cite{Astaneh:2014uba} and \cite{W2}. Referring to this definition, we can find the decomposition energy for an interval of length $\ell$ on a two dimensional CFT which simply reads $\CE=\frac{1}{\ell^2}$.

Moving away from the conditions that provide such a simple structure for the EE naturally complicates the computations. So we do not expect to have this simple expansion for any modification of a field theory. Nevertheless, there are some exceptions and an important example is the $T\bar{T}$ deformation of a CFT. Such a deformation is very special in the sense that it provides an example of an irrelevant deformation of a CFT which is still UV-complete. It is this integrability while being irrelevantly deformed that makes this theory so special \cite{Zamolodchikov:2004ce}, \cite{Smirnov:2016lqw} and \cite{Cavaglia:2016oda}.

This deformation is known through the following equation
\be
\delta_\mu S_{QFT}=\int d^2x\, O_{T\bT}\, ,
\ee
where $\mu$ is the parameter of deformation and
\be
O_{T\bT}=\Tr T^2-(\Tr T)^2\, .
\ee
Interestingly, there are two dominant rich holographic pictures in correspondence with this deformation. One picture comes from string theory on some UV/IR interpolating backgrounds \cite{Asrat:2017tzd}-\cite{Giveon:2017nie} and according to the other recipe, which is called the finite cut off prescription, the deformation is equivalent to removing the asymptotic region of AdS$_3$. We pursue the latter prescription by putting the theory on a finite cut off boundary of AdS. This finite radius is closely bound up with the deformation parameter $\mu$. In this sense $\mu$ plays the role of a geometrical cut off \cite{Hi}, see also \cite{Guica}.

This holographic picture enables us to define some sort of $T^2$ deformation in the higher dimensions. For a $d$ dimensional CFT, the composite operator is defined by the following form \cite{Taylor}, (see also \cite{Cardy1} for a different generalization to the higher dimensions)
\be
O_{T^2}=\Tr T^2-\frac{1}{d-1}(\Tr T)^2\, .
\ee
This geometrical picture affects the expressions we obtained for quantum mechanical quantities including the holographic EE.
In this note, we seek to show how the EE changes due to this deformation. Furthermore, we want to scan the whole spectrum of the deformed theory. 
An interesting result that we will state in advance is that there are some pseudo thermodynamical laws for this change.

The paper is organized as follows: In section 2 we propose our set of laws of entanglement thermodynamics for $T^2$-deformed theories. In section 3 we compute the EE for excited states of $T\bar{T}$-deformed theory using a direct holographic calculation which verifies our second law. In the rest of that section, we present a method based on the calculation of the holographic minimal area to extend this analysis to the higher dimensions. The final section is devoted to conclusions.
\section{Entanglement Laws of $T^2$-deformations}
Suppose that $T^2$-deformed theory lives on a $d$-dimensional manifold, $\CM$, covered by coordinates $x^i$ and with metric $g_{ij}$ on it. We assume that this base manifold is located at a finite radius of an asymptotically AdS spacetime $\mathcal{N}$, constructed using the Fefferman-Graham expansion The coordinates in the bulk space would be $X^\mu=\{\rho,x^i\}$ and we denote the metric on $\mathcal{N}$ by $G_{\mu\nu}$. In this setup the finite (here non-zero) radius is of the same dimension as the deformation parameter $\mu$ and it can be placed in an explicit and exact correspondence as $\rho_c=\frac{\mu N}{16\pi}$.

We assume a time constant slice of $\CM$ and decompose it to two subregions by means of a spacelike surface of co-dimension two, $\Sigma$. The coordinates and the induced metric on $\Sigma$ will be denoted by $y^a$ and $h_{ab}$.

The EE would then be the von-Neumann entropy for the reduced density matrix for each subsystem which is enclosed by $\Sigma$
\be
S_{EE}=-\Tr\, \varrho\log\varrho\, .
\ee
According to \emph{Ryu-Takayanagi}'s proposal, this quantity could be calculated holographically. To do so, one needs to minimally extend the entangling surface $\Sigma$ into the bulk to obtain a closed minimal holographic hypersurface, $\Sigma_H$. Let us denote the coordinates and the induced metric on $\Sigma_H$ by $Y^A=\{\rho,y^a\}$ and $H_{AB}$, respectively. Then the EE would be simply obtained in terms of the area of this surface \cite{Ryu:2006bv}, \cite{Ryu:2006ef}
\be
S_{EE}=\frac{\CA_{\Sigma_H}}{4G_N}\, .
\ee
The profile of the minimal surface will be determined as a solution of the following \emph{Gauss-Weingarten} equation, see \eg \cite{Poisson}
\be
\p_a\p^a X^i+\Gamma[G]^i_{jk}\p_aX^j\p^aX^k-h^{ab}\Gamma[H]^c_{ab}\p_c X^i-\Gamma[H]^\rho_{ab}\p_\rho X^i=0\, .
\ee
The minimal surface meets the entangling surface at the location of the finite cut off and thus one may consider the following expansion for that
\be
X^{i}(x,\rho)=X^{(0)i}+(\rho-\rho_c)X^{(1)i}+\cdots\, .
\ee
Putting this into the Gauss-Weingarten equation and imposing the minimality condition we get at the leading order of $\rho$
\be
X^{(1)i}=-\frac{K n^i}{2(d-2)}\, ,
\ee
where $n^i$ and $K$ are the unit normal and the trace of the extrinsic curvature on $\Sigma$. The induced metric on $\Sigma_H$ reads
\be
H_{AB}=\p_A X^\mu \p_B X^\nu G_{\mu\nu}\, ,
\ee
in which the leading terms read
\be\label{H1}
\begin{split}
&H_{\rho\rho}=\frac{L^2}{4\rho^2}\left[1+\rho\frac{K^2}{(d-2)^2}+\CO(\rho^2)\right]\ ,\\
&H_{ab}=\frac{L^2}{\rho}\left[h_{ab}-(\rho-\rho_c)\frac{KK_{ab}}{(d-2)}+\CO(\rho^2)\right]\ .
\end{split}
\ee
Now we need to construct the area element, $\sqrt{H}$ and calculate the area of the minimal surface.
Doing so, we get an expansion for the EE for a theory defined on a lattice with characteristic length $\rho_c$. 
The result obtained in this way has an interesting and significant interpretation. First, it should be noted that the deformation parameter $\mu$ defines a limit above which, some sorts of non-locality becomes apparent. However we do not exceed this limit and thus the leading term still admits an area law. Interestingly enough, although the deformed theory is a complicated operator theory in comparison with the undeformed CFT, the structure of the subleading term also remains unchanged. Of course, the numerical prefactors will be significantly deformed. Putting the original CFT on the same lattice on which the deformed theory lives, the change of the entanglement entropy admits the following laws

\textbf{First Law of $T^2$ deformation}
\\
The difference of entanglement entropies of the modified theory and the original theory satisfies a first law
\be
\delta S_\Sigma=\delta\CE_\Sigma=\frac{1}{\tau}\CE_\Sigma\ \ , \ \ \tau=\frac{\rho_c^{-2+\frac{d}{2}}}{N}\ .
\ee
This means that the change in EE in a deformed theory, which is expected to be very complex, is entirely provided by the bending energy of the entangling surface.\\
Note that we have defined a \emph{deformation temperature}, $\tau$ which contains the parameter of deformation in it\footnote{ In four dimensions the dependence of the sub-leading term of the EE to the cutoff (and thus to the parameter of deformation) becomes logarithmic. This is a well-known \emph{universal} feature of the bending energy in four dimensions. This universality can be understood in this way that one can always arbitrarily add a constant logarithm, $\log\lambda$ to scale the parameter of deformation in logarithmic dependence as $\log\lambda\mu$. This means that the coefficient of the bending energy in four dimensions tells us only about some fundamental characteristics of the theory such as the effective numbers of the fields. In order to probe the details of the entanglement thermodynamics of deformation, one needs to look at the higher moments of the bending energy of the entangling surface which is proportional to $\int_\Sigma K^4$. We leave the study of different aspects of this interesting case to future works. We would like to acknowledge the interesting comment by the referee on this point.}.

\textbf{Second Law of $T^2$ deformation}
\\
The above law is a saturation limit for a general inequality that is satisfied along the spectrum. In fact, the excitation energy provides a portion of the energy of deformation and thus in general we have
\be
\delta S_\Sigma\leq \frac{1}{\tau}\CE_\Sigma\, .
\ee
As we will see later, this law is a restatement of the preservation of the unitarity bound.

\textbf{Third Law of $T^2$ deformation}
\\
Since the deformation temperature, $\tau$ is defined in terms of a finite cut off radius, it is a positive quantity which never vanishes nor diverges.

In particular the first law can be verified in $d=2$ for $T\bar{T}$-deformation, in which
\be\label{two}
\tau=\frac{1}{N\rho_c}\rightarrow \frac{1}{\tau}\CE_\Sigma=\frac{\mu c^2}{72\pi\ell^2}\, ,
\ee
that is exactly the modification of the EE which we get due to the $T\bar{T}$-deformation. This verifies our proposal that the change of EE for this deformation equals to a specific change in the energy of decomposition. In the next section we will prove the second law for this deformation in two dimensions. In this way, we also discover a first law in the heart of the first law of $T\bar{T}$- deformation.
\section{Probing the whole spectrum of $T\bar{T}$-deformed theory}
We begin our study of the whole spectrum by considering a two-dimensional deformed theory which lives on a cylinder of unit radius covered by, $x=\sigma+it$ and its conjugate. We want to compute the EE for a subsystem $\sigma\in I=[a,b]$ of length $\ell=b-a$ on the $t=0$ slice of the cylinder. Then one needs to compute the reduced density operator for this subsystem, $\varrho$ and calculate the von-Neumann entropy

In a field theory, $\varrho$ is governed by partial tracing over the complementary part of the interval. Denoting the typical field content of the theory by $\psi$ and the action by $S[\psi]$ we evaluate
\be
\varrho_{+-}=\frac{1}{Z_1}\int\CD\psi\, e^{-S[\psi]}\Big\vert_{\substack{\psi(\sigma_I,0^-)=\psi_-\\ \psi(\sigma_I,0^+)=\psi_+}}\, .
\ee
Thus the reduced density operator will be given by a path integral along the cylinder by two cuts at the location of the entangling interval. $Z_1=\Tr\varrho$ is the partition function which will be obtained by identifying the cuts.
 It is not easy to directly calculate the von-Neumann entropy in field theories so we usually change the order of logarithm and trace by introducing the Renyi entropy
\be
S_n=\frac{1}{1-n}\log\Tr\varrho^n\, . 
\ee
Then it is easy to check that $S_{EE}=\lim_{n\rightarrow 1}S_n$.
This quantity can be evaluated following the replica trick. Doing so, we prepare $n$ copies of the theories on $n$ cylinders which on each there is a cut along the entangling interval. Then the three steps of defining the reduced density operator, constructing its $n$'th power and performing the trace are realized through the evaluation of the partition function on a $n$-Cylinder, $\CC_n$  which is constructed by gluing the cylinders along the cuts. 
\be
Z_n=\int_{\CC_n} D\psi\, e^{-S[\psi]}.
\ee
Here $Z_n$ is the partition function on $\CC_n$.
Then
\be
S_n=\frac{1}{1-n}\log\frac{Z_n}{Z_1^n}\, .
\ee
The details of calculation can be found in \cite{Calabrese:2004eu} and \cite{Calabrese:2009qy}.

In order to scan the whole spectrum of the theory we insert incoming and outgoing operators at infinite past and future times, respectively, $t=\mp\infty$.
\be
(\varrho_O)_{+-}=\frac{1}{Z_1}\lim_{x\rightarrow i\infty}\frac{1}{\la O(x,\bx) O^\dagger(-x,-\bx)\ra}\int \CD\psi\, e^{-S[\psi]}O(x,\bx)O^*(-x,-\bx)\Big\vert_{\substack{\psi(\sigma_I,0^-)=\psi_-\\ \psi(\sigma_I,0^+)=\psi_+}}\, .
\ee
 Then the main challenging problem would be the calculation of the partition function on a replicated geometry in the presence of the insertions. This issue was addressed for the first time in \cite{Berganza:2011mh} and \cite{Alcaraz:2011tn} and then in \cite{Mosaffa} with a different method. The latter one uses the symmetric orbifolding which was introduced in \cite{Lunin:2000yv}.

 We perform a series of transformations to make this calculation possible. In one line, these transformations are as follows
\be
\CC_n(x,\bx)\xrightarrow[T_1]{w=e^{-ix}}\CR_n(w,\bw)\xrightarrow[T_2]{z=\left(\frac{w-u}{w-v}\right)^\frac{1}{n}}\BC(z,\bz)\xrightarrow[T_3]{y=-i\log z}\CC(y,\by)
\ee
The first transformation, $T_1$ moves us to the sheets instead of the cylinders. This enables us to make the calculations more easily. These sheets are sewn together to construct the so called $n$-sheeted Riemann surface, $\CR_n$.
The metric on $\mathcal{R}_n$ reads
	\be
	ds^2=dw d\bw\, .
	\ee
	Despite the simple form of the metric the geometry on $\CR_n$ suffers from conical singularities at the location of the cuts. What we need to compute on this manifold is 
	\begin{equation}
\Tr\varrho^n_O\equiv\frac{Z_n}{Z_1^n}\lim_{\substack{w\rightarrow 0\\ w'\rightarrow \infty}}
\frac{\langle \displaystyle \prod_{k=0}^{n-1}O_k(w,\bar{w})
O_k^\dagger(w',\bar{w'}) \rangle_{\mathcal{R}_n}}{\langle
O(w,\bar{w})O^\dagger(w',\bar{w'})\rangle^n_{w}}\ ,\label{corrcylinder}
\end{equation}
	but obviously, it is not a straightforward calculation due to the singularities which arise in the geometry as a result of replication.
	 This motivates us to perform the second transformation, $T_2$ in order to regularize the singularity, \cite{Calabrese:2004eu}
	\be\label{wtoz}
	z=\left(\frac{w-u}{w-v}\right)^{1/n}\, .
	\ee
	Then 
	\be
	ds^2=\left\vert \frac{\p w}{\p z}\right\vert^2 dz d\bz=e^{\xi(z,\bz)}dzd\bz\, ,
	\ee
	where
	\be
	e^{\frac{\xi(z,\bz)}{2}}=n\vert u-v\vert\frac{\vert z\vert^{n-1}}{\vert z^n-1\vert^2}\, ,
	\ee
is a conformal factor in which all the complexities caused by replication are encoded. In this way, we move to a complex plane $\mathbb{C}$ which is covered by coordinates $(z,\bz)$. In fact, performing the transformation $T_2$, the edges of the interval, map to origin and infinity, respectively and each sheet forms a wedge. Then by putting these wedges together, a complex plane is made.
	On this complex plane, the insertion points lie on a circumference of a circle of unit radius\footnote{The insertion points are in fact the infinite past and infinite future which correspond to $w\rightarrow -i\infty$ and $w\rightarrow +i\infty$, respectively. Moving to the complex plane, it is the coordinate transformation \eqref{wtoz} that tells us what distribution the insertion points would find. Remembering that $u=e^{-ia}$ and $v=e^{-ib}$ the past/future insertion points would be found as the $n$'th root of $e^{i(b-a)}$ and $1$, respectively, \ie\,  $z_{k,n}=e^{\frac{2i\pi}{n}(\theta+k)}$ and $z'_{k,n}=e^{\frac{2i\pi}{n}k}$. In order to display these insertion points in a more symmetrical way, we apply a phase shift in the amount of $e^{-\frac{i\pi}{n}\theta}$. Doing so we arrive at equation \eqref{insertions}. Obviously, because in all functions and transformations, the coordinates are associated with their conjugates, this phase shift does not change the final real results. This phase shift was originally introduced in references \cite{Berganza:2011mh} and \cite{Alcaraz:2011tn}.}
\be\label{insertions}
z_{k,n}=e^{\frac{i\pi}{n}(\theta+2k)}\ \ , \ \ z'_{k,n}=e^{\frac{i\pi}{n}(-\theta+2k)}\ \ , \ \ k=0,1,\cdots\, , \ \ \theta=\frac{\ell}{2\pi}\, .
\ee
As a result, the two-point correlation function of the insertions on the replicated geometry is transformed into the $2n$-point correlation functions on a complex plane up to a conformal factor, $T_O$. 
\be
{\CF_n}\vert_{({\text{insertion points}})}=\lim\limits_{\substack{w\rightarrow0\\
w'\rightarrow\infty}}\frac{\langle \displaystyle \prod_{k=0}^{n-1}O_k(w,\bar{w})
O_k^\dagger(w',\bar{w}') \rangle_{R_n}}{\langle
O(w,\bar{w})O^\dagger(w,\bar{w}')\rangle^n_{w}}=T_O
\frac{\langle \displaystyle
\prod_{k=0}^{n-1}O(z_{k,n},\bar{z}_{k,n})O^\dagger(z'_{k,n},\bar{z}'_{k,n})
\rangle_{\BC}}{\langle O(z_{0,1},\bar{z}_{0,1})
O^\dagger(z'_{0,1},\bar{z'}_{0,1})\rangle^n_{\BC}}\ .\ee
The last transformation brings us to a regular cylinder of unit radius
\be
\CF_n(\theta)=T_O
\frac{\langle \displaystyle
\prod_{k=0}^{n-1}O(y_{k,n},\bar{y}_{k,n})O^\dagger(y'_{k,n},\bar{y}'_{k,n})
\rangle_{\CC}}{\langle O(y_{0,1},\bar{y}_{0,1})
O^\dagger(y'_{0,1},\bar{y'}_{0,1})\rangle^n_{\CC}}\ ,
\ee
where
\be
y_{k,n}=\frac{\pi}{n}(\theta+2k)\ \ , \ \ y'_{k,n}=\frac{\pi}{n}(-\theta+2k)\ \ , \ \ k=0,1,\cdots\, .
\ee
But even on a smooth regular cylinder, this calculation is not straightforward at all. 
In addition to the complexities of calculating the $T_O$ coefficient, the changes of the OPEs and correlation functions due to the $T\bar{T}$-deformation are very non-trivial and difficult to perform, see \cite{Cardy:2019qao} for instance. As always, holography cleans the scene and simplifies the calculation by introducing some geometrical substitutions. This is what we do in the following. 

	We perform a holographic calculation in order to calculate the transformation factor $T_O$ for CFT$_2$ deformed by $T\bar{T}$ deformation. The usual way to do this calculation is to use the Field-Operator map. Assuming a scalar theory in AdS$_{d+1}$ one may draw a clear correspondence with an operator theory on CFT side. The massive scalar theory is introduced with the following simple action
	\be
	I=-\frac{1}{16\pi G_N}\int d^{d+1}X\sqrt{G}(\p_\mu \Phi\p^\mu\Phi+M^2\Phi^2)\, .
	\ee
The equation of motion for the scalar field gives two independent solutions which could be identified with their asymptotic behavior. Denoting the radial coordinate in the bulk with $\rho$, the asymptotic form of the scalar field takes the following form
\be
\Phi(\rho,x)\sim \phi_0(x)\rho^{\frac{\Delta_-}{2}}+\phi_1(x)\rho^{\frac{\Delta_+}{2}}\, ,
\ee
where
\be
\Delta_{\pm}=\frac{d}{2}\pm\sqrt{\frac{d^2}{4}+M^2}\, ,
\ee
are two distinct roots of the equation $\Delta(\Delta-d)=M^2$ \cite{Witten:1998qj}, \cite{Aharony:1999ti}.

In this expansion, $\phi_1(x)$ gives the vacuum expectation value of the dual operator $O(x)$ with conformal dimension $\Delta_+$ and $\phi_0$ stands as a source for that.\footnote{Henceforth, we set a common notation in which $\Delta_+=\Delta$ and $\Delta_-=d-\Delta=\alpha$.}
So referring to the AdS/CFT correspondence
\be
Z_{SUGRA}[\phi_0]=e^{W_{CFT}[\phi_0]}=\la e^{-\int_{\mathbb{C}}\phi_0 O}\ra_{CFT}\, ,
\ee
where $Z_{SUGRA}[\phi_0]$ and $W[\phi_0]$ are the partition function and the generating function in gravity and CFT sides at the asymptotic limit.

This correspondence allows us to obtain the n-point functions of the operators on the CFT side with the help of variation of the normalized action of gravity plus matter with respect to the source \cite{deHaro:2000vlm}, \cite{Skenderis:2002wp}. To be precise, it can be stated that
\be
\la O(x)\ra_{CFT}=\lim_{\epsilon\rightarrow 0}\left(\frac{1}{\epsilon^\Delta\sqrt{h}}\frac{\delta I_{ren}[\Phi]}{\delta\phi_0(x)}\vert_{\phi_0=0}\right)\, ,
\ee
where $\epsilon$ and $h$ are the parameter of the UV cut off and the metric of the asymptotic boundary, respectively.
This is exactly what we need to do in order to calculate the EE. In fact, such a correspondence simplifies the computation of CFT correlators on a complicated geometry in terms of some simpler variations of the bulk action. This calculation is presented in full detail below.

The metric of AdS$_3$ in FG gauge takes the following form 
	\be
	ds^2=\frac{L^2}{4\rho^2}d\rho^2+\frac{L^2}{\rho}g_{ij}(\rho,w)dw^idw^j\ \ \ , \ \ \  g_{ij}(\rho,w)=g^{(0)}_{ij}(w)+\rho g^{(1)}_{ij}(w)+\cdots\, ,
	\ee
	where $w^{1,2}=(w,\bw)$ and $g^{(0)}_{ij}dw^idw^j$ will be identified with the metric on $n$-sheeted Riemann surface. 
Interestingly enough, there is a set of coordinate transformations in AdS$_3$
	which brings us to Poincare form of the metric \cite{Krasnov}, \cite{Myers1}. 
To put it bluntly, by applying the following transformations
\be\label{FG to Poincare}
	r=\frac{\rho^{1/2}e^{-\xi/2}}{1+\frac{1}{4}\rho e^{-\xi}\vert\p_w\xi\vert^2}\ \ , \ \ z=w+\frac{1}{2}\, \frac{\rho e^{-\xi}\bp_w\xi}{1+\frac{1}{4}\rho e^{-\xi}\vert\p_w\xi\vert^2}\, .
\ee
one arrives at
\be\label{PAdS}
	ds^2=\frac{L^2}{r^2}(dr^2+dzd\bz)\, .
\ee
As a result of this coordinate transformation, all calculations can be done in Poincare coordinates, with the stipulation that the cut off surface is a non-constant surface of radius one, and all of the non-trivialities of the complex replicated geometry are reflected in the profile of that surface. Denoting the radius of the cut off surface in Poincare coordinates by $r_c$, one finds that
 up to the order $\CO(\rho^{1/2})$, the coordinate transformation \eqref{FG to Poincare} yields $w=z$ and
	\be\label{cutoff}
	r_c^{-1}=\rho_c^{-1/2}e^{\frac{\xi(z,\bz)}{2}}+\frac{1}{4}\rho_c^{1/2} e^{-\frac{\xi(z,\bz)}{2}}\Xi(z,\bz)\, ,
	\ee
	where
	\be
	\Xi(z,\bz)=\p\xi(z,\bz)\bar{\p}\xi(z,\bz)\, .
	\ee
Evaluating the gravitational on-shell action in this geometry with non constant cut off we get the EE for the ground state of a $T\bar{T}$-deformed theory, \eqref{two}. This is the subject of \cite{Astaneh1}. Here we want to calculate the EE for the whole spectrum of the deformed theory. To do so, we need to turn a scalar on in the bulk. 
Let us denote the scalar field in Fefferman-Graham and Poincare coordinates by $\Phi$ and $\Phi'$, respectively. One clearly knows that
\be
\Phi(\rho,w,\bw)=\Phi'(r,z,\bz)\, .
\ee
The scalar field admits an asymptotic expansion in radial coordinates
\be
\begin{split}
&\Phi(\rho,w,\bw)=\rho^{\frac{\alpha}{2}}\sum_{n=0}\rho^n\phi_{2n}(w,\bw)\, ,\\
&\Phi'(r,z,\bz)=r^{\alpha}\sum_{n=0}r^{2n}\phi'_{2n}(z,\bz)\, .
\end{split}
\ee
which on the finite radius cut off yields
\be
\phi'_{2n}=\rho_c^{\frac{2(n+1)-\Delta}{2}}r_c^{-2(n+1)+\Delta}\phi_{2n}\, .
\ee
This expansion brings us to an important relation between the vacuum expectation values in replicated and regular spaces
\be
\la O(w,\bw)\ra_{\mathcal{R}_n}=\rho_c^{-\frac{\Delta}{2}}r_c^{\Delta}\la O(z,\bz)\ra_{\mathbb{C}}\, .
\ee
Referring to the source-response correspondence we observe that moving from the complex plane to the replicated geometry can be restated as the rescaling of the scalar source
\be
\phi_0\rightarrow \rho_c^{\frac{\Delta}{2}}r_c^{-\Delta}\phi_0\, .
\ee
This is a very important observation, which reduces all computational complexities on a replicated geometry and for the deformed theory to an appropriate redefinition of the source. Using this, one immediately concludes that
\be
T_O=
\frac{\displaystyle\prod_{k=0}^{n-1}\rho_c^{-\frac{\Delta}{2}}r_c^\Delta(z_{k,n},\bz_{k,n};z'_{k,n},\bz'_{k,n})}{\left(\vert(z_{0,1}-1)(z_{0,1}'-1)\vert/\vert u-v\vert\right)^{2n\Delta}}\, ,
\ee
in which
\be
\begin{split}
&e^{-\frac{\xi(z_{0,1},\bz_{0,1})}{2}}=\frac{1}{n}\tan(\frac{\pi\theta}{2})\, ,\\
&\Xi(z_{0,1},\bz_{0,1})=(1-n^2)+n^2\csc^2(\frac{\pi\theta}{2})\, ,\\
&\frac{\vert (z_{0,1}-1)(z_{0,1}'-1)\vert}{\vert u-v\vert}=\tan(\frac{\pi\theta}{2})\, ,
\end{split}
\ee
and similar relations hold at $(z'_{k,n},\bz'_{k,n})$.\\
By putting everything in its place, the leading terms of the conformal factor $T_O$ are found as follows
\be
T_O=n^{-2n\Delta}\left\{1-\frac{\rho_c\Delta}{2n}\left[n^2-1+\sec^2(\frac{\pi\theta}{2})\right]\right\}\, .
\ee
Therefor, the factor $\CF_{n}$ gets the following correction due to placing the boundary at a finite cut off
\be
\begin{split}
\delta\CF_n(\theta)&=-\frac{\rho_c\Delta}{2n}n^{-2n\Delta}\left[n^2-1+\sec^2(\frac{\pi\theta}{2})\right]\times\frac{\langle \displaystyle
\prod_{k=0}^{n-1}O(y_{k,n},\bar{y}_{k,n})O^\dagger(y'_{k,n},\bar{y}'_{k,n})
\rangle_{\CC}}{\langle O(y_{0,1},\bar{y}_{0,1})
O^\dagger(y'_{0,1},\bar{y'}_{0,1})\rangle^n_{\CC}}\ .
\end{split}
\ee
Note that for small entanglement intervals, a very good approximation can be found for the remainder of the expression. In fact using the OPE one deduces that
\be
\frac{\langle \displaystyle
\prod_{k=0}^{n-1}O(y_{k,n},\bar{y}_{k,n})O^\dagger(y'_{k,n},\bar{y}'_{k,n})
\rangle_{\CC}}{\langle O(y_{0,1},\bar{y}_{0,1})
O^\dagger(y'_{0,1},\bar{y'}_{0,1})\rangle^n_{\CC}}=n^{2n\Delta}(1+\frac{1-n^2}{3n}\Delta\pi^2\theta^2)+\cdots\, ,
\ee
where dots includes the suppressed contribution from the primaries other than the identity operator.

\be
\overset{(O)}{\delta} S_\Sigma=-\p_n \CF_n(\theta)\vert_{n=1}=\frac{1}{6}\Delta\ell^2-\frac{(3+8\Delta)\Delta}{96}\rho_c\ell^2+\cdots\, .
\ee
Here we have explicitly marked the variation by writing a superscript to emphasize that it is the change of the EE for excited states of a deformed theory. The first term, which is found for the undeformed theory has been produced holographically in \cite{Astaneh2}. The second term is what we get now for the deformed theory. Clearly, this contribution is negative which guarantees that our second law of deformation is valid. Moreover, here we recognize a first law in the heart of the previous first law of deformation. To see this, we note that the above structure can be considered as the first few terms of the following closed form
\be
\overset{(O)}{\delta} S_\Sigma=\frac{\ell^2}{3\tilde{\rho}_c}\left(1-\sqrt{1-\tilde{\rho}_c\Delta}\right) \ \ , \ \ \tilde{\rho}_c=-\frac{3+8\Delta}{4\Delta}\rho_c\, ,
\ee
which, according to the identification $\rho_c\sim\mu$, evokes the structure of the energy spectrum of a $T\bar{T}$ deformed theory, see \eg \cite{Bonelli:2018kik}.

Now let us make a comment on higher dimensional cases.
In the dimensions higher than two, it is practically not possible to perform a direct holographic calculation of the EE similar to what we did in two dimensions. But it is still possible to provide a specific analysis of the pattern of entanglement for excited states of the $T^2$-deformed theory,  based on the minimal surface prescription. Doing this, what we need to take into account is the change of the background geometry due to the external scalar field. Solving the Einstein equations in Fefferman-Graham coordinates and in the presence of a source, we find that the metric on the asymptotic boundary gets the following correction \cite{Blanco:2013joa}-\cite{Casini:2016rwj}
\be
g_{ij}=\eta_{ij}+\delta g_{ij}\ \ , \ \ \delta g_{ij}=m\rho^{\beta}\eta_{ij}\, ,
\ee
where $\eta_{ij}$ is the flat metric on the base manifold and $\beta$ represents the different possible powers in the expansion of the deformed metric. These are $d/2$, $\Delta$ and $d-\Delta$ for which the mass scale $m$ is proportional to $\phi_0\phi_1$, $\phi_0^2$ and $\phi_1^2$, respectively. The mass scale is negative provided by the unitarity bound. We have had a clear intention that we did not write the explicit form of the constant prefactor. In fact, by solving the Einstein equations only the trace and the covariant divergence of such terms would be determined. As a result, it is always possible to shift these coefficients in the metric with some freedom. However, what is important for us is the sign of this term, which is determined by the unitarity bound.

Turning a scalar on in the bulk, the induced metric on $\Sigma_H$ changes as follows
\be\label{H2}
\begin{split}
&\delta H_{\rho\rho}=\frac{mL^2K^2}{4(d-2)^2}\rho^{\beta-1}\ ,\\
&\delta H_{ab}=\left[mL^2h_{ab}-\frac{m L^2KK_{ab}}{(d-2)}(\rho-\rho_c)\right]\rho^{\beta-1}\ .
\end{split}
\ee
These changes have been found by solving the Gauss-Weingarten equation in which the deformed metric has been put in. One should note that the extremization is achieved by zeroing the external field and due to this fact the backreaction of the metric is the only source for the change in the profile of the minimal surface. So all that we need to evaluate is $\delta\sqrt{H}=\frac{1}{2}\sqrt{H}(H^{AB}\delta H_{AB})$. Then one should perform an integral over $\rho$ which goes from $\rho_c$ to $\rho_*$, where $\rho_*$ is the point at which the volume element vanishes. Doing so one finds
\be\label{HD}
\overset{(O)}{\delta} S_\Sigma=\frac{(d-2)N^{\frac{d}{2}}}{2[2(\beta+1)-d]}\int_\Sigma m\rho^{\beta+1-\frac{d}{2}}\big\vert_{\rho_c}^{\rho_*}\, .
\ee
Interestingly enough, if we impose the unitarity bound on each of values of $\beta$ we get a positive denominator and positive power of $\rho$
\be
\begin{split}
&\beta=\frac{d}{2}\ \rightarrow \ (\beta+1)-\frac{d}{2}=1\, ,\\
&\beta=\Delta\ \rightarrow \ (\beta+1)-\frac{d}{2}=(\Delta-\frac{d-2}{2})>0\, ,\\
&\beta=d-\Delta\ \rightarrow \ (\beta+1)-\frac{d}{2}=(\frac{d}{2}-\Delta+1)>0\, ,\\
\end{split}
\ee
and thus the whole contribution is negative which verifies our second law of deformation. For this reason, we stated earlier that our second law is the statement of establishing the unitarity bound.

Another interesting aspect of this equation is the behavior in two dimensions. Although, according to what we mentioned before, it is impossible to determine the exact value of the numerical coefficients in this analysis, the dependence on the length of the entanglement interval can be extracted. First of all, it should be noted that in leading order $\rho_*=\frac{(d-2)}{K^2}$, which for a sphere of radius $R$ equals to $\frac{R^{4-d}}{(d-2)}$. Note that we have considered the contribution of the fields to the stress tensor and thus we have chosen $\beta=d/2$. So the factor $(d-2)$ in \eqref{HD} drops and a dependence of the form of $-\ell^2$ comes out in two dimensions which is exactly what we have produced in the direct holographic calculation. This ensures us again that our set of laws of deformation are valid in various dimensions.
\section{Conclusion}
In this article, we have investigated the pattern of entanglement in a $T^2$-deformed theory and in particular in a two dimensional CFT deformed by $T\bar{T}$-insertion. Based on our findings, although at first glance it may seem that the quantum features of the deformed theory have significant differences compared to the original one, these changes can be included in simple and intuitive geometrical concepts. Taking this idea seriously, we have arrived at a set of very simple and comprehensive laws for the changes of the EE that are very similar to the laws of thermodynamics.

But why do we think these laws are intuitive and natural?
The $T\bar{T}$ and its generalization to the higher dimensions, $T^2$-deformed theories, are all about a certain composite operator constructed by the stress tensor whose expectation value is surprisingly factorized at not only special limits.
Referring to existing holographic prescriptions, the deformed theory lives on a finite cut off boundary of AdS spacetime. Then, the factorization of the composite operator could be translated to the Gauss-Codazzi equation for the Brown-York energy on the finite cut off radius which is written in terms of the extrinsic curvature of the boundary. This identity states that on the flat boundary of AdS$_{d+1}$
\be
\CK^2-\CK_{ij}\CK^{ij}=\frac{d(d-1)}{L^2}\, ,
\ee
where $\CK_{ij}$ is the extrinsic curvature tensor on the finite cut off boundary. This relation helps us to eliminate $\Tr \CK^2$ in favour of $\CK^2$ to finally prove that the Brown-York stress tensor factorizes and its determinant (in the contexts of QFT$_2$/AdS$_3$) equals its trace up to some numerical prefactors. Relying on the power and capabilities of holography, this observation can be generalized to higher dimensions as well.\\
In order to calculate the EE we consider a time constant slice of this boundary. We assume time translational symmetry which results that the extrinsic curvature tensor is vanishing in this direction. So finally we are left with the integral of the square of the trace of extrinsic curvature in a spacial direction. This is exactly the bending energy of the entangling surface which is written in terms of the Willmore's functional of the surface. So we guessed that the change of the entropy due to the $T^2$-deformation in general, should be related to the change of the bending energy of the entangling surface. Refinement of this conjecture led us to the definition of energy of decomposition and the first law of $T^2$-deformation. We have explicitly shown that this first law confirms the previous results in two dimensions for $T\bar{T}$-deformed CFT.

As explained in details in the text, a second law also appears which tells us about the establishment of the unitarity bound. According to this law, all the coefficients go hand in hand to beautifully convince us that the excitation energy supplies some amount of the bending energy and therefore introduces an inequality whose saturation limit gives the first law.

The relationship between energy and entropy in the first law leads us to define the deformation temperature in terms of the finite radius cut off. Naturally, this temperature never vanishes nor diverges. This is the subject of the third law of deformation.

We should note that there may still be other equalities and inequalities between various quantities related to EE and geometrical characteristics of the entangling surface that would complete our list of laws. One can also test our laws for other deformations like that $J\bar{T} (\bar{J}T)$ which are the result of introducing a new global $U(1)$ current \cite{Chakraborty:2019mdf}. The non-relativistic limits of deformed theories have been investigated in some articles, see for instance \cite{Alishahiha:2019lng} and \cite{Jeong:2022jmp}. Our laws of entanglement thermodynamics can be tested for them as well. It would also be very interesting to search whether similar sets of laws exist for other quantities in quantum information theory, \eg entanglement negativity, quantum complexity etc. We leave all these interesting problems for future works.
\section*{Acknowledgment}
We would like to thank Horacio Casini for useful discussions. We would also like to thank Alireza Hassanzadeh for working on the early stages of this work.

\end{document}